\documentclass[preprints,article,accept,moreauthors,pdftex]{Definitions/mdpi}

\usepackage{bm}
\usepackage{bbm}
\usepackage{listings}
\usepackage{subfig}
\usepackage{amsmath}
\usepackage{amssymb}
\usepackage{booktabs}
\usepackage{mathtools}
\usepackage{hyperref}
\usepackage{fourier}
\usepackage[T1]{fontenc}
\usepackage[]{avant}
\usepackage{longtable}

\firstpage{1} 
\pubvolume{}
\issuenum{1}
\articlenumber{1}
\pubyear{2024}
\copyrightyear{2024}
\history{}

\NewDocumentCommand{\expect}{ e{^} s o >{\SplitArgument{1}{|}}m }{%
  \operatorname{E}
  \IfValueT{#1}{{\!}^{#1}}
  \IfBooleanTF{#2}{
    \expectarg*{\expectvar#4}%
  }{
    \IfNoValueTF{#3}{
      \expectarg{\expectvar#4}%
    }{
      \expectarg[#3]{\expectvar#4}%
    }%
  }%
}
\NewDocumentCommand{\expectvar}{mm}{%
  #1\IfValueT{#2}{\nonscript\;\delimsize\vert\nonscript\;#2}%
}
\DeclarePairedDelimiterX{\expectarg}[1]{[}{]}{#1}

\newcommand*\diff{\mathop{}\!\kern0pt\mathrm{d}}

\Title{Counterexamples for FX Options Interpolations - Part I}
\Author{Jherek Healy}
\AuthorNames{Jherek Healy}

\address{jherekhealy@protonmail.com}
\abstract{This article provides a list of counterexamples, where some of the popular fx option interpolations  break down. Interpolation of FX option prices (or equivalently volatilities), is key to risk-manage not only vanilla FX option books, but also more exotic derivatives which are typically valued with local volatility or local stochastic volatilility models.}

\begin{document}
	\section{Introduction}
Market prices of Foreign exchange (FX) options are typically quoted as a sparse set volatilities per option maturity.
Those correspond to the volatilities of an at-the-money (ATM) option, 25$\Delta$  and 10$\Delta$ risk-reversals and butterflies.
Depending on the currency pair and the maturity, there is much variation in what ATM exactly means: is it at the money straddle? with or without premium? spot or forward ? There are also two conventions for the risk-reversal (RR) and butterfly (BF): the simple smile convention where \begin{align*}
\sigma_{\textmd{call}} = \sigma_{\textmd{ATM}} + \sigma_{\textmd{BF}} + \frac{1}{2}\sigma_{\textmd{RR}}\,, \quad \sigma_{\textmd{put}} = \sigma_{\textmd{ATM}} + \sigma_{\textmd{BF}} - \frac{1}{2}\sigma_{\textmd{RR}}\,,
\end{align*} and the more involved broker convention which requires a numerical solver.  \citet{clark2011foreign,reiswich2012fx} describe
the various conventions in details. 

The conversion of RR and BF quotes to vanilla option implied volatilities will be the focus of a subsequent article. In this paper, the concern is to price and risk manage options, whose strikes will not correspond to the strikes implied
by the quoted $\Delta$. A popular approach is to interpolate the volatilities in terms of $\Delta$. A single $\Delta$ convention
is typically used for all maturities, regardless of the actual quoting $\Delta$ convention.	
\citet{malz1997option} proposed to use a quadratic polynomial, cubic splines are another common choice.
Interpolating in $\Delta$, requires to solve the following non linear problem in order to find the volatility for a given option strike:
\begin{equation}
  \sigma(K) = f\left( \Delta_F(K,\sigma(K))\right)\,,\label{eqn:volforstrike}
\end{equation}
where $f$ is the smile representation, $\Phi$ the cumulative normal distribution, $\sigma$, the implied volatility, and \begin{equation}
\Delta_F(K,\sigma(K)) =\pm \Phi\left(\pm \frac{\ln \frac{F(T)}{K} - \frac{\sigma^2(K)T}{2}}{\sigma(K)\sqrt{T}}
\right)\,,\end{equation} for the forward delta convention without premium (+ for a call, - for a put). In \citep{malz1997option} the function $f$ is a quadratic, while in \citep{clark2018using} it is the exponential of a quartic.

To avoid the non-linear problem, \citet{clark2011foreign} suggests the use of a polynomial in terms of a reduced $\Delta_R$.
\begin{equation}
  \Delta_R(K) = \Phi\left(\frac{\ln \frac{F(T)}{K}}{\sigma(F(T))\sqrt{T}}\right)\,.\label{eqn:delta_reduced_clark}
\end{equation}
In the expression of $\Delta_R$, the volatility is taken at-the-money only.

Those choices are simple and the representation in $\Delta$ is convenient for the traders, but they suffer from two major
drawbacks: they are not arbitrage-free, and they lead to a flat volatility for small and large strikes.
Flattening wings are not particularly realistic, as this means that all moments are finite \citep{lee2004moment}, and thus
 the distribution does not really have fat tails. In addition, stochastic volatility models lead to linear wings in variance \citep{gatheral2011convergence}.

Some practitioners thus moved to more advanced representations, such as the SABR model of \citet{hagan2002managing} (with $\beta=1$),
or the SVI parameterization of \citet{gatheral2006volatility}. Those do not allow
to interpolate exactly the option quotes, due to the constrained shapes of volatility smiles they can represent.
One on hand it may avoid overfitting, on the other hand, it is sometimes be desirable to fit nearly exactly to the market quotes.
For example, Clark uses the quartic representation to be able to detect some multi-modality in the implied probability density. Other uses
may relate to some risk engine limitations.

The key characteristics of an interpolation scheme for fx option volatilities are:
\begin{enumerate}
	\item no arbitrage,
	\item smooth implied probability density,
	\item ability to control the wings,
	\item possibility of nearly exact interpolation, even if this usually means overfitting the quotes.
\end{enumerate}

In this article, we will provide a list of counterexamples, where some of the popular smile representation break down.
\section{Oscillations in the density}
\begin{figure}[H]
	\centering{
		\includegraphics[width=\textwidth]{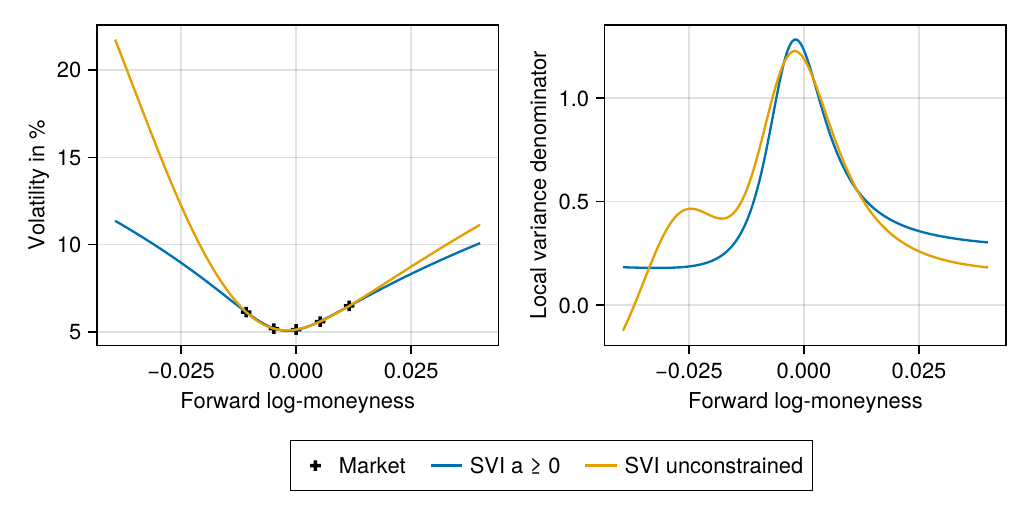}}
	\caption{Smile and local variance denominator for SVI calibrated to AUD/NZD options expiring in 7 days of of 2014/07/02.\label{fig:audnzd_svi_denom}}
\end{figure}

\citet{wystup2017arbitrage} presented one of the rare counterexample of the literature to illustrate the possibility of
negative density with the SVI parameterization, and provided detailed market data, which allow for an easy replication of the issue. 
It consists of FX options on AUD/NZD expiring in 7 days as of 2014/07/02. The quotes, reproduced in Table \ref{tbl:audnzd_7d} thus correspond to the convention of spot $\Delta$ with premium.
\begin{table}[H]
	\caption{Options on AUD/NZD expiring in 7 days as of 2014/07/02. $F(T)=1.07845$, $B_{\textmd{AUD}}(T)=0.999712587139$, spot = 1.0784. The quotes are converted to Put/Call vanilla vols using the simple approximation.\label{tbl:audnzd_7d}}
	\centering{\begin{tabular}{ccccc}\toprule
	ATM &25$\Delta$-RR & 25$\Delta$-BF & 10$\Delta$-RR & 10$\Delta$-BF \\
	5.14 & 0.40 & 0.25 & 0.35 & 1.175  \\\cmidrule(lr){1-5}
		10$\Delta$-Put& 25$\Delta$-Put & ATM & 25$\Delta$-Call & 10$\Delta$-Call \\
	6.14 & 5.19 & 5.14 & 5.59 & 6.49 \\\bottomrule
	\end{tabular}}
\end{table}
If we calibrate SVI according to the recommendations of \citet{de2009quasi}, that is enforcing $a \geq 0$, and thus ensuring that the variance is always positive, the 
density stays positive. In Figure \ref{fig:audnzd_svi_denom}, we plot the Dupire local variance denominator \cite{gatheral2006volatility} \begin{equation}
  g(y) = 1 - \frac{y}{w(y)} \frac{\partial w}{\partial y} + \frac{1}{4} \left(-\frac{1}{4}- \frac{1}{w(y)} + \frac{y^2}{w(y)^2}\right) \left(\frac{\partial w}{\partial y}\right)^2 + \frac{1}{2}\frac{\partial^2 w}{\partial y^2}
\end{equation} 
where $w(y)=\sigma^2(F(T)e^{y}) T$ is the total variance in log-moneyness, $F(T)$ is the forward price to maturity $T$. In the SVI representation, we have $w(y) = aT + bT (\rho (y-m) +\sqrt{(y-m)^2+ s^2}$ where $a,b,\rho,m,s$ are SVI parameters).
A negative variance denominator is equivalent to a negative probability density, but the scale of the denominator function makes it easier to visualize the location of negative density. 
While the constraint works well on this example, it is not guaranteed to always work, although, in general, the SVI parameterization behaves well, hence its popularity.

\begin{figure}[H]
	\centering{
		\includegraphics[width=\textwidth]{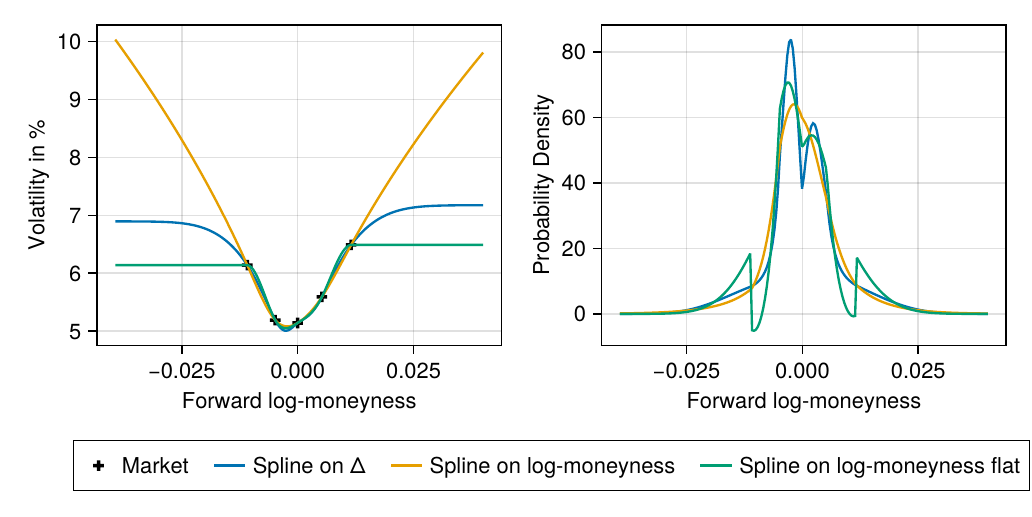}}
	\caption{Smile and probability density for cubic spline in log-moneyness or $\Delta$ calibrated to AUD/NZD options expiring in 7 days of of 2014/07/02.\label{fig:audnzd_spline_dens}}
\end{figure}

It turns out that when the volatilities are interpolated with a cubic spline on $\Delta_F$, a 
spurious spike appears spline density near the money. Although
\citet[Figure 1]{wystup2017arbitrage} mentions this problem for another example, it is much more significant in Figure \ref{fig:audnzd_spline_dens}, possibly
because of the choice of interpolation axis. When applying the cubic spline to log-moneyness $y$ and variances $\sigma^2(y)$, we end up only with
a slight bimodality, more in-line with \citep[Figure 1]{wystup2017arbitrage}. 
The boundary conditions used for the spline interpolation also play a key role: if we set the first derivative to be zero at the boundaries and use a flat extrapolation, the probability density presents two spurious spikes and goes negative, due to the discontinuities of the second derivative at the $10\Delta$-Put and Call. If, instead, we set the second derivative to be zero at the boundaries (also called natural cubic spline), along with a linear extrapolation of the same slope, the second derivative is continuous everywhere.

Even when the probability density stays positive, spikes at the extrapolation points are undesirable: the numerical $\Gamma$ of vanilla options will present the same spikes and thus leads to vastly inaccurate hedges. In Figure \ref{fig:audnzd_spline_dens}, we also notice that two modes clearly appear in the density, when using the flat extrapolation and a log-moneyness axis, which were almost invisible with the linear extrapolation and a log-moneyness axis. This is another side-effect of the spline boundary conditions.

\begin{figure}[H]
	\centering{
		\includegraphics[width=\textwidth]{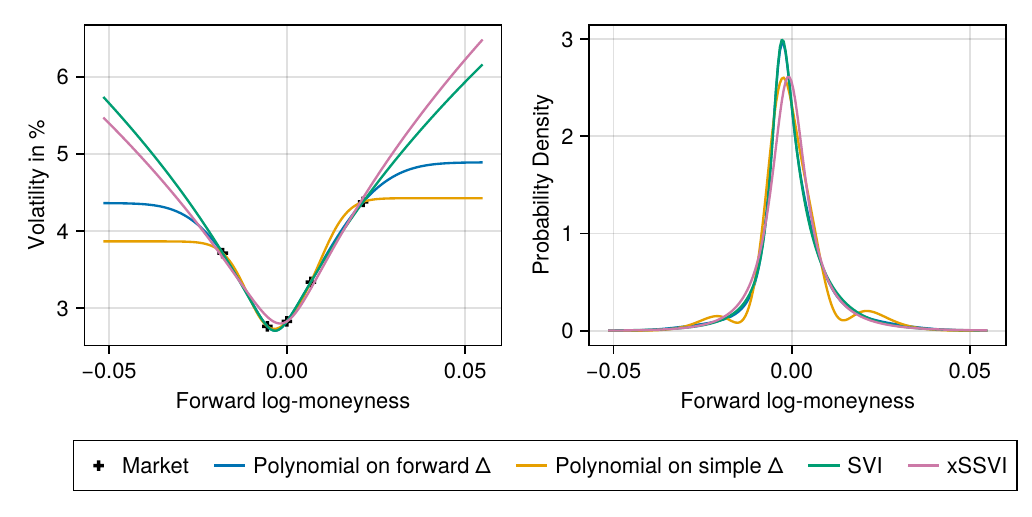}}
	\caption{Smile and probability density for polynomial in simple or forward $\Delta$ calibrated to EUR/CZK options expiring in 32 days of of  2019/12/16.\label{fig:eurczk_poly_dens}}
\end{figure}

\citet{clark2011foreign} suggests to use a polynomial in simple $\Delta_R$, as it then does not require the solution of a non-linear problem to lookup the volatility for a given strike.
This is a priori attractive, but it also leads to oscillations. 
The oscillations are not limited to this example, but are present for many short term options cases. Figure \ref{fig:eurczk_poly_dens} shows the implied
probability density for options\footnote{While unusual, the use of 5-$\Delta$ is sometimes the practice. Here the quotes are directly those of put and call options and may have been calibrated in an external system.} on EUR/CZK (Euro vs. Czech Koruna) with 32 days to maturity as of 2019/12/16 (Table \ref{tbl:eurczk_32d}). 
A remedy is to consider a polynomial in $\Delta_F$ and solve the non-linear problem.

\begin{table}[H]
	\caption{Options on EUR/CZK expiring in 32 days as of 2019/12/16 in the forward $\Delta$ with premium convention. $F(T)=1.07845$, $B_{\textmd{AUD}}(T)=0.999712587139$, spot = 1.0784. \label{tbl:eurczk_32d}}
	\centering{\begin{tabular}{ccccc}\toprule
			5$\Delta$-Put& 25$\Delta$-Put & ATM & 25$\Delta$-Call & 5$\Delta$-Call \\
			3.715 &2.765 &2.830 &3.340 &4.380\\\bottomrule
	\end{tabular}}
\end{table}


\section{Low volatilities}
The AED currency (United Arab Emirates Dirham) is pegged to the USD, and exhibits extremely low volatility as a consequence. Surprisingly, there are quotes for FX options on USD/AED. Table \ref{tbl:usd_aed_9m} presents quotes for two maturities as of 2023/01/24. The problems are similar for the different maturities, and we will consider options expiring in 9 months to illustrate the issues that arise with different interpolations.
	
\begin{table}[H]
	\caption{Quotes for options on USD/AED as of 2023/01/24 in forward $\Delta$ with premium convention. For the 9m maturity, we have $F(T)=3.67206$, $r_{\textmd{USD}}(T)=3.255\%$, spot = 3.67. The quotes are converted to Put/Call vanilla vols using the simple approximation.\label{tbl:usd_aed_9m}}
	\centering{\begin{tabular}{cccccc}\toprule
	Maturity & ATM &25$\Delta$-RR & 25$\Delta$-BF & 10$\Delta$-RR & 10$\Delta$-BF \\
	1m & 0.31 & 0.142 & 0.078 & 0.343 &0.142\\
9m &	0.32 & 0.152 & 0.084 & 0.412 & 0.392  \\
1y & 0.29 & 0.132 & 0.072 &0.359 & 0.343 \\ \cmidrule(lr){2-6}
		& 10$\Delta$-Put& 25$\Delta$-Put & ATM & 25$\Delta$-Call & 10$\Delta$-Call \\
	9m &0.506 & 0.328 & 0.32 & 0.48 & 0.918 \\\bottomrule
	\end{tabular}}
\end{table}

\begin{figure}[H]
	\centering{
		\includegraphics[width=\textwidth]{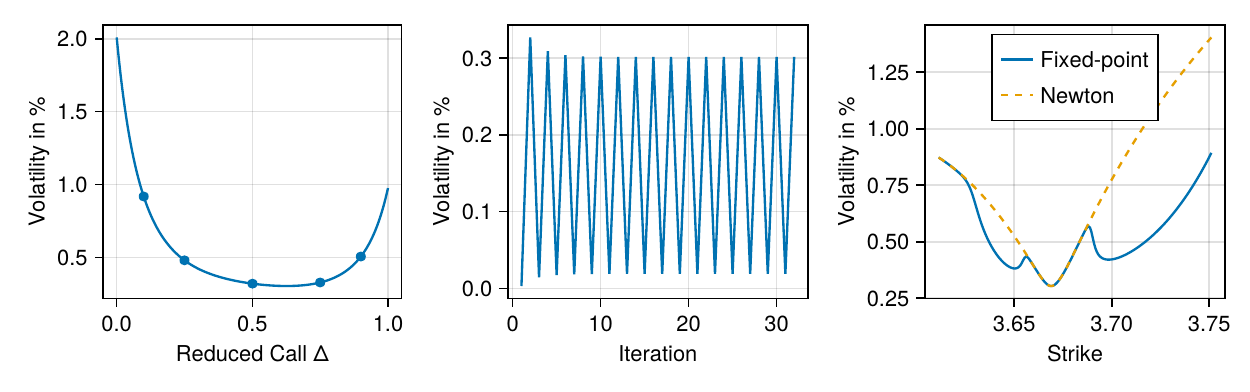}}
	\caption{Smile and convergence of volatility lookup at strike $K=3.7$ of the exponential quartic interpolation for options on USD/AED options expiring in 9 months as of 2023/01/24.\label{fig:usdaed_poly_conv}}
\end{figure}

If we fit the exponential polynomial function in $\bar{\Delta}_{F}$ to those quotes, where \begin{equation}\bar\Delta_F = \Phi\left(\frac{1}{\sigma\sqrt{T}}\ln (F(T)/K)\right)\label{eqn:bardeltaf}\,,\end{equation}
 we need to be careful how to solve the non-linear Equation \ref{eqn:volforstrike} or equivalently Equation \ref{eqn:bardeltaf}. In particular, the fixed point method described in \citep[Section 1.4.9]{wystup2017fx} does not converge. For a strike price of 3.7, the implied $\bar\Delta_F$ oscillates between two points: 1.9\% and 30.1\%, neither are a solution. The implied volatility oscillates as well, Figure \ref{fig:usdaed_poly_conv} shows the first 32 iterations, increasing the number of iterations does not help. Because of the failure to lookup properly the volatility at a given strike, and we stop the fixed point method at a finite number of iterations, the implied volatility obtained by the fixed point method has a strange shape. Newton or Brent methods work well.  We employ $\bar\Delta_F$ to make the analysis simpler, but the same observations would hold with $\Delta_F$.

\begin{figure}[H]
	\centering{
		\includegraphics[width=\textwidth]{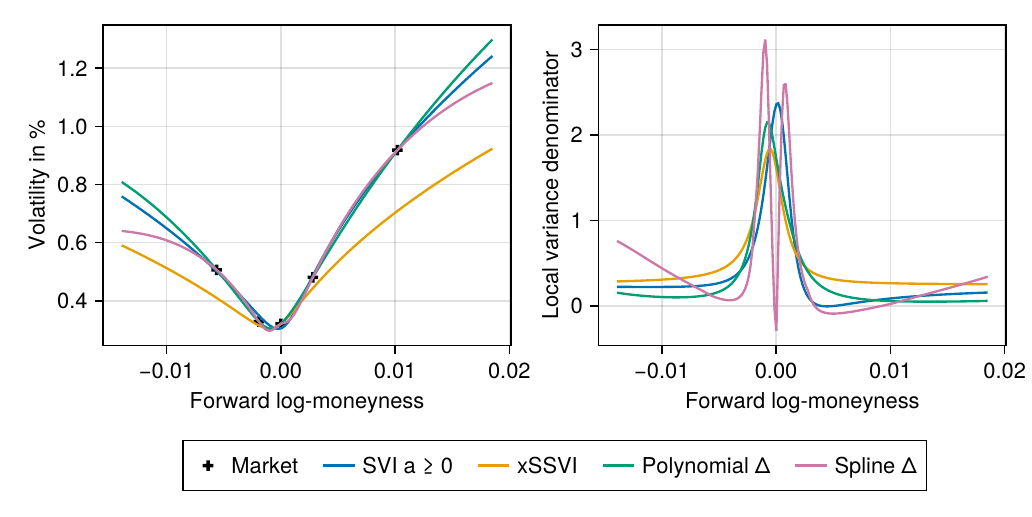}}
	\caption{Smile and local variance denominator with different interpolations of options on USD/AED options expiring in 9 months as of 2023/01/24.\label{fig:usdaed_svi_dens}}
\end{figure}

This example also illustrates a negative density with SVI and the constraint\footnote{Without the constraint, SVI would actually lead to arbitrage free prices on this example. Yes, it is the reverse as for AUD/NZD.} $a \geq 0$ (Figure \ref{fig:usdaed_svi_dens}). 
The extended SSVI (xSSVI) smile is not matching the 25$\Delta$ quotes properly, due to its limited number of parameters and additional constraints that makes the local variance denominator positive. We used the calibration procedure described in \citep{corbetta2019robust}. 
The exponential quartic polynomial on $\bar\Delta_F$ works well when Newton's method is used. SABR also provides an excellent fit (better than SVI), with a positive density.
The natural cubic spline in $\Delta_F$ with linear extrapolation leads to negative density at the money.


\section{Negative volatilities}
We consider now options on EUR/TRY of maturity 1 year as of 2022/11/29. The Turkish Lira is particularly volatile, due to the high inflation in Turkey in 2022 and the resulting implied volatility smile has a somewhat unusual shape. We calibrated the vanilla options volatilities using the exponential quartic function on $\bar\Delta_F$ (Table \ref{tbl:eurtry_1y}) and fit various smile representations on those vanilla options.

\begin{figure}[H]
	\centering{
		\includegraphics[width=\textwidth]{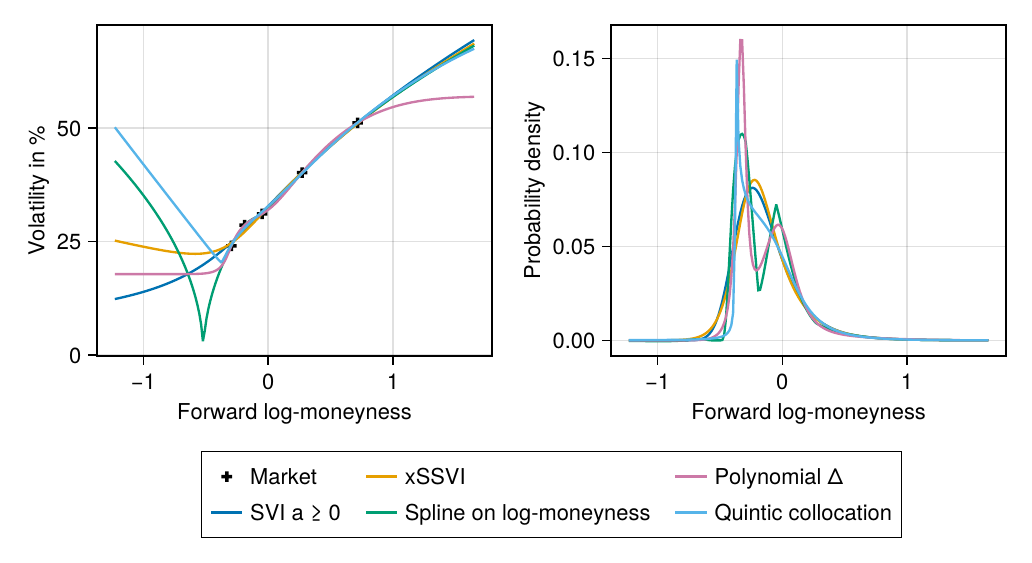}}
	\caption{Smile and probability density with different interpolations of options on EUR/TRY options expiring in 1 year as of 2022/11/29.\label{fig:eurtry_svi_dens}}
\end{figure}

\begin{table}[H]
	\caption{Quotes for options on EUR/TRY as of 2022/11/29 in forward $\Delta$ with premium. For the maturity of 184 days, we have $r_{\textmd{TRY}} = 36.77\%$, $r_{\textmd{EUR}}=1.167\%$ . For the 1 year maturity, we have $r_{\textmd{TRY}} = 37.73\%$, $r_{\textmd{EUR}}=1.784\%$, spot=19.3483.\label{tbl:eurtry_1y}}
	\centering{\begin{tabular}{cccccc}\toprule
	Maturity & ATM &25$\Delta$-RR & 25$\Delta$-BF & 10$\Delta$-RR & 10$\Delta$-BF \\
	6m & 22.12 & 9.385 & 2.187 & 21.148 & 7.633\\
1y	& 31.13 & 11.568 & 2.931 & 27.120 & 9.307 \\\cmidrule(lr){2-6}
	&	10$\Delta$-Put& 25$\Delta$-Put & ATM & 25$\Delta$-Call & 10$\Delta$-Call \\
1y	& 24.08 & 28.64 & 31.13 & 40.21 & 51.20 \\\bottomrule
	\end{tabular}}
\end{table}
The exponential polynomial representation suggests a bimodal density (Figure \ref{fig:eurtry_svi_dens}). This results in a spike in the probability density of the quintic collocation of \citet{lefloch2019model1} (which otherwise behaves well on the other examples considered in this paper) and a corresponding unnatural steep angle in the implied volatilities. SVI and xSSVI do not fit exactly the quotes, but are close enough while providing a smooth unimodal probability density. SABR (not displayed) leads to a similar fit as xSSVI.

It is a priori not obvious that the probability density is truly bimodal, or if bimodality is a consequence of overfitting. In this case, it is likely the latter. More problematic is the natural cubic spline on log-moneyness, which produces negative implied variance, because the slope at the 10$\Delta$ is positive.

\section{Manufactured example}
It is possible to find more awkward examples that break many of the smile parameterization by generating
distributions from a mixture of two lognormal distributions with random\footnote{This means 4 random parameters: one weight, one mean, and two volatilities. The sum of weights equals 1 and global mean equals the forward.} parameters, calculating implied volatilities for given $\Delta_F$, until 
some of the parameterization misbehaves. Such an example is provided in Table \ref{tbl:man_1y}, for a maturity of 1 year,  domestic and foreign rates equal to zero, and a spot price of 1.0. We use a forward $\Delta$ convention without premium. We corrected it a little bit to make it more interesting: the 10$\Delta$ quotes are a flat extrapolation of the 25$\Delta$ quotes.
\begin{table}[H]
	\caption{Manufactured option quotes of maturity 1 year. $r_d = 0\%$, $r_f=0\%$, spot=1.0.\label{tbl:man_1y}}
	\centering{\begin{tabular}{ccccc}\toprule
	10$\Delta$-Put& 25$\Delta$-Put & ATM & 25$\Delta$-Call & 10$\Delta$-Call \\
	26.00 & 26.00& 19.50 & 12.70 & 12.70 \\\bottomrule
	\end{tabular}}
\end{table}

\begin{figure}[H]
	\centering{
		\includegraphics[width=\textwidth]{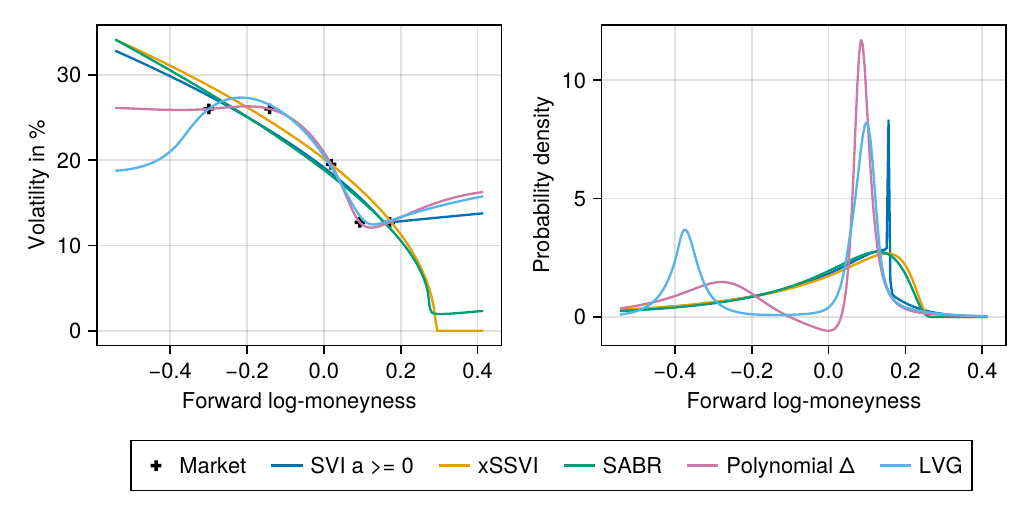}}
	\caption{Smile and probability density with different interpolations of manufactured options quotes.\label{fig:man_svi_dens}}
\end{figure}

SABR and xSSVI do not fit well but leads to similarly reasonable smiles given the awkward quotes. The SVI smile looks more odd, with a strong angle. A remedy would be to increase its $s$ parameter. The exponential polynomial on $\Delta$ leads to a negative density (Figure \ref{fig:man_svi_dens}). Cubic splines in log-moneyness or $\Delta_F$ (not displayed) also result in a negative density. Finally, the quadratic local variance gamma (LVG) of \citet{lefloch2023quadratic} allows to fit the quotes very well, with a positive density. 

We stress that the original quotes do not have any arbitrage. They may serve as a warning against flat extrapolation, and especially against adding fictitious flatly extrapolated quotes. The latter often leads to a multi-modal implied probability density.

\section{Wings control}
Table \ref{tbl:eurusd_1m} presents the implied volatilities for a dense range of vanilla put and call forward $\Delta$ for options on EURUSD of maturity 1 month as of 2022/03/11, as may be retrieved from a market data provider such as Reuters or SuperDerivatives \citep[Section 1.5.9]{wystup2017fx}.
\begin{table}[H]	
	\caption{EURUSD option vols of maturity 1 month as of 2022/03/11 in forward $\Delta$ without premium convention. $F(T) = 0.975848$, $r_{\textmd{USD}} = 0.351836\%$, spot=0.9759.\label{tbl:eurusd_1m}}
	\centering
	\resizebox{\textwidth}{!}{%
	\begin{tabular}{@{}ccccccccccccccccccc@{}}\toprule
		1 & 5 & 10 & 15 & 20&  25& 30& 35& 40&  (P) ATM (C) & 40 & 35& 30& 25&  20 & 15& 10& 5& 1\\
		14.04 & 13.01& 12.47& 12.16 & 11.93 & 11.73 & 11.54 & 11.38 & 11.25 & 11.02 & 10.85 & 10.78 & 10.73 & 10.68 & 10.63 & 10.58 & 10.51 & 10.45 & 11.11 \\\bottomrule
	\end{tabular}}
\end{table}

\begin{figure}[H]
	\centering{
		\includegraphics[width=\textwidth]{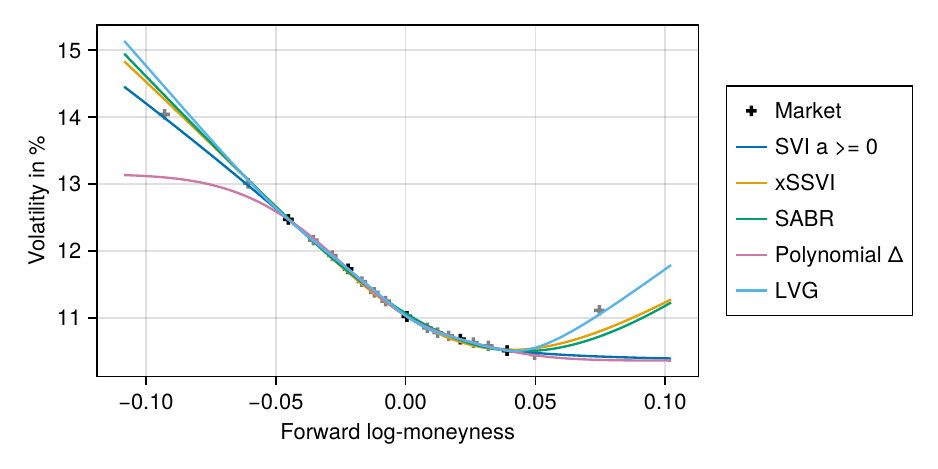}}
	\caption{Smile with different interpolations fitted to the 5 usual options quotes of maturity 1 month, for EURUSD  as of as of 2022/03/11.\label{fig:eurusd_extra}}
\end{figure}

We fit the various smile representations to the 5 usual option quotes (ATM, 25$\Delta$ and 10$\Delta$). They have no problem matching those 5 quotes, but their wings differ significantly. Figure \ref{fig:eurusd_extra} shows that the extrapolation with $\Delta$ based interpolations such as the exponential polynomial in $\bar\Delta_F$ looks too flat compared to what is quoted by the provider. The left (put) wing of SVI looks good, but the right (call) wing is not as good, although it concerns only one observation: the 1$\Delta$-Call, which may not be very liquid. SABR and xSVVI are
 very similar and have reasonable left and right wings, while LVG leads to a good fit of the right and left wings. 
\begin{figure}[H]
	\centering{
		\includegraphics[width=\textwidth]{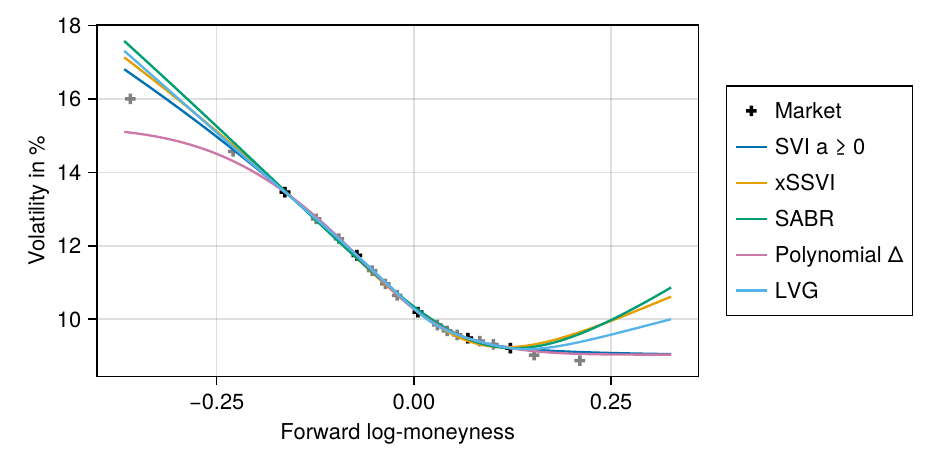}}
	\caption{Smile with different interpolations fitted to the 5 usual options quotes of maturity 1 year, for EURUSD  as of as of 2022/03/11.\label{fig:eurusd_extra_1y}}
\end{figure} 

It is enlightening to look at the case of options maturing in one year (Table \ref{tbl:eurusd_1y} and Figure \ref{fig:eurusd_extra_1y}). The exponential polynomial  in $\Delta$ looks better, although still too flat in the left wing and SVI is best.
\begin{table}[H]	
	\caption{EURUSD option vols of maturity 1 year as of 2022/03/11 in forward $\Delta$ without premium convention. $F(T) = 0.986772$, $B_{\textmd{USD}}(T) = 0.963113\%$, spot=0.9759.\label{tbl:eurusd_1y}}
	\centering
	\resizebox{\textwidth}{!}{%
	\begin{tabular}{@{}ccccccccccccccccccc@{}}\toprule
		1 & 5 & 10 & 15 & 20&  25& 30& 35& 40&  (P) ATM (C) & 40 & 35& 30& 25&  20 & 15& 10& 5& 1\\
		16.0 & 14.57 & 13.46 & 12.73 & 12.19 & 11.74& 11.32& 10.96& 10.65& 10.19& 9.84& 9.68& 9.57& 9.48& 9.4& 9.31& 9.21& 9.02& 8.87\\
		\bottomrule
	\end{tabular}}
\end{table}

\citet{wystup2023slope} explains how much extrapolation can play a role in the prices of digital options. There are other important basic use cases where extrapolation play a role: auto-quantos and variance swaps. An auto-quanto option on EUR/USD is the same as a regular vanilla option on EUR/USD except for the payment currency: the auto-quanto pays out the amount $\max\left(\pm(S(T)-K),0\right)$ in EUR (foreign currency), for a call (resp. a put). This is equivalent to an option on  $\max\left(\pm(S(T)-K),0\right)S(T)$ in USD (domestic currency). The latter can be priced by replication of vanilla options \citep[Section 7.5]{castagna2010fx}:
\begin{align}
	C_{\textmd{auto-quanto}}(K,T) = 2 \int_K^\infty C(K,T) \diff K + K C(K,T)\,,\quad P_{\textmd{auto-quanto}}(K,T) = -2 \int_0^K P(K,T) \diff K + K P(K,T)\,,
\end{align}
where $C(K,T)$ and $P(K,T)$ are call and put options of maturity $T$ and strike $K$.
Due to the flat wings, the $\Delta$ based smiles will undervalue the auto-quanto call options and overvalue the put. For the EUR/USD options with maturity 1 year, the impact is around 3\% of the option price for strikes at $10\Delta$ (Table \ref{tbl:autoquanto_eurusd}), with a larger discrepancy for the 
put. The discrepancy is of course much more significant if we are in the extrapolation part, for example for strikes at $1\Delta$, it is a factor of 2: even larger than digitals.

The price of a newly issued variance swap, which uses a replication of vanilla options across all strikes may also be impacted. 
If we use the approach of \citet{doi:10.1142/S0219024911006681}, and only the 5 reference quotes, we end up with a model independent price of 11.28 (in terms of vol\footnote{The present value of the variance swap contract is the square of the vol price.}) for the 1 month and 11.20 for the 1 year maturity. This is quite close to the prices obtained by replicating on a dense range of options with strikes spanning over 5 standard deviations from the forward, with the various smile representations (Table \ref{tbl:autoquanto_eurusd}). The smile representation has a negligible impact for the 1 month maturity, and reaches up to 3\% of the variance swap present value for the 1 year maturity. This due to the representation of the left wing, which plays a more important role than the right wing for variance swaps, due to the $1/K^2$ weighting in the replication.
\begin{table}[H]
	\caption{Prices derivative contracts on EUR/USD for a notional of 10,000 as of 2022/03/11, using different smile representations. For the 1-month maturity, the $10\Delta$ Put and Call strikes are $K_{10,P} = 0.93274$, $K_{10,C}=1.014727$ and the $1\Delta$ strikes are $K_{1,P} = 0.889339$, $K_{1,C}=1.051435$.\label{tbl:autoquanto_eurusd}}
	\centering
\begin{tabular}{lllrrrrr}\toprule
Maturity & Option type & Strike &Polynomial $\Delta$ & SVI & xSSVI & SABR & LVG \\\midrule
1m & Digital Put &$K_{10,P}$ &1063.88 & 1064.91 & 1067.09 & 1066.47 & 1063.88\\
& &$K_{1,P}$ & 70.76 & 108.69 & 122.34 & 125.44 & 132.10 \\
& Auto-quanto Call& $K_{10,C}$ & 14.28 & 14.34 & 14.37  & 14.31&  14.33 \\
& & $K_{1,C}$ & 0.60 & 0.63 & 0.86 & 0.81 & 1.04\\
&Auto-quanto Put &$K_{10,P}$  & 15.12 & 15.07 & 15.09 & 15.07 & 15.00\\
& &$K_{1,P}$  & 0.66 & 1.13 & 1.30 & 1.34 & 1.43\\
&Variance swap & 0 & 11.31 & 11.35 & 11.37 & 11.39 & 11.41 \\
1y& Digital Put& $K_{10,P}$ & 1210.64 & 1211.74 & 1215.64 & 1269.08 & 1210.64\\
 & & $K_{1,P}$ & 101.21 & 185.30 & 204.72 & 232.28 & 232.82\\
&Auto-quanto Call& $K_{10,C}$ & 47.86 & 48.24 & 49.06  & 48.49&  48.17 \\
&& $K_{1,C}$ & 4.21&4.39&6.97 & 6.80 & 5.60 \\
&Auto-quanto Put &$K_{10,P}$  & 46.39 & 44.92 & 44.72 & 44.14 & 44.20\\
& &$K_{1,P}$  & 2.14 & 4.27 & 4.80 & 5.47 & 4.97\\
&Variance swap & 0&10.89 & 11.03 & 11.07 & 11.16 & 11.10 \\
\bottomrule
\end{tabular}
\end{table}

\section{Conclusion}
Using a cubic spline is dangerous, natural boundaries and log-moneyness axis should be favored. Beside the possibility of negative or oscillating implied probability density, the natural cubic spline in log-moneyness suffers from the more important issue of a possible positive slope for the put wing extrapolation, which will need to be dealt with in a likely awkward way. 
Interpolations in $\Delta$ space lead in general to too flat wings, and may misprice auto-quanto options and variance swaps of long maturities, as well as digital options beyond the $10\Delta$ strikes. Another take-away, if you were to use a $\Delta$ based representation nonetheless is: don't use the fixed-point method to lookup the volatility at a given strike.

While SVI makes sense for FX smiles, the occasional SVI arbitrages still need to be dealt with, possibly with a recalibration with a penalty. In this regard, xSSVI is attractive but may fit poorly $10\Delta$ quotes in some cases. SABR was found to be similar to xSSVI in terms of fit and did not lead to negative density issues on the examples considered, but it is not guaranteed to be arbitrage-free, especially at the surface level, across maturities.
The quadratic local variance gamma model was found to provide a flexible arbitrage-free interpolation, at the cost of a more complex implementation.

In this article, the smile representations were fitted directly to vanilla option quotes, we will explore the role and impact of interpolation during the calibration to broker quotes in a follow-up article.

\acknowledgments{The author would like to thank Dr. Gary Kennedy and Dr. Fabien Le Floc'h for fruitful conversations and feedback on early drafts of this paper.}
\externalbibliography{yes}
\bibliography{arbfree_interpolation.bib}
\appendixtitles{no}

\appendix
	
\end{document}